\documentclass[useAMS,usenatbib]{mn2e}
\usepackage{graphicx}
\usepackage{txfonts}
\usepackage{natbib}
\newcommand{\xmm}{{\it XMM-Newton}}

\newcommand{\einstein}{{\it Einstein Observatory}}

\def\gsim{\mathrel{\hbox{\rlap{\hbox{\lower4pt\hbox{$\sim$}}}\hbox{$>$}}}}
\def\lsim{\mathrel{\hbox{\rlap{\hbox{\lower4pt\hbox{$\sim$}}}\hbox{$<$}}}}

\begin{document}
   \title[]{The linear rms-flux relation in an Ultraluminous X-ray Source}

   \author[L.M. Heil, S. Vaughan]{L.M. Heil and S. Vaughan \\
   X-Ray and Observational Astronomy Group, University of
   Leicester,  Leicester, LE1 7RH, U.K.\\}

   \date{Draft \today}

   \pagerange{\pageref{firstpage}--\pageref{lastpage}} \pubyear{2002}

   \maketitle
   
   \label{firstpage}

   \begin{abstract}
	We report the first detection of a linear correlation between rms
	variability amplitude and flux in the Ultraluminous
	 X-ray source NGC 5408 X-1. The rms-flux relation has
	previously been observed in several Galactic black hole X-ray binaries
	(BHBs), several Active Galactic Nuclei (AGN) and at least one neutron
	star X-ray binary. This result supports the
	hypothesis that a linear rms-flux relation is common to all luminous
	black hole accretion and perhaps even a fundamental property of
	accretion flows about compact objects. We also show for the first time the cross-spectral properties of the
variability of this ULX, comparing variations below and above $1$ keV. The
coherence and time delays are poorly constrained but consistent with high
coherence between the two bands, over most of the observable frequency
range, and a significant time delay (with hard leading soft variations). The magnitude and frequency dependence of the
 lags are broadly consistent with those commonly observed in BHBs, but the
 direction of the lag is reversed. These results indicate that ULX variability
 studies, using long X-ray observations, hold great promise for constraining
 the processes driving ULXs behaviour, and the position of ULXs in the scheme
 of black hole accretion from BHBs to AGN. 

   \end{abstract}

 

\section{Introduction}
\label{sect:intro}
	X-ray variability studies have played a crucial role in furthering our understanding of accreting black hole systems. Among the more recent discoveries in this area was the linear relationship between the rms amplitude of the variability and its flux, a relation which appears to persist over a wide range of timescales. To date this has been observed of four black hole X-ray binaries (BHBs) and one neutron star X-ray binary \citep{Uttley01, Gleissner04, Uttley05, Gandhi09} and several Active Galaxies \citep{Uttley01, Vaughan03b, Vaughan03a}. If present in accretion around stellar mass ($\sim$10 M$_{\odot}$) and supermassive black holes ($>$ 10$^{6}$ M$_{\odot}$) it seems reasonable to hypothesise that the linear rms-flux relation is common to accreting black holes of all masses. In this case we would expect such a relation to be present in Ultraluminous X-ray sources (ULXs). 

	ULXs were first discovered by the \einstein~ \citep{Fabbiano89} as point-like sources with luminosities greater than 10$^{39}$ erg~ s$^{-1}$. Timing and energy spectral studies suggest they are similar to accreting compact objects such as black hole binaries (BHBs) and active galactic nuclei (AGN). The inferred isotropic luminosities often exceed the Eddington limit for a stellar mass black hole ($ < 20~M_{\odot}$). Consequently it has been suggested that this, along with their low inferred disc temperatures \citep{Miller04}, indicates that they represent a new class of intermediate mass black holes with masses ranging from 100-1000 M$_{\odot}$. Counter to this are theories that assume a lower mass black hole (e.g $<$ 100 $M_{\odot}$) accreting at a super-Eddington rate \citep{Begelman02}, emitting anisotropically \citep{King01} or through relativistic beaming \citep{Georgan02}.

	X-ray timing studies of bright ULXs have shown some to display timing properties common to BHBs such as spectral breaks and quasi-periodic oscillations (QPOs) \citep{Strohmayer03, Strohmayer07, Strohmayer09, Heil09}. However, it has also been suggested that both energy spectral and timing behaviour within ULXs may differ from those seen from BHBs and AGN \citep{Stobbart06, Goad06, Gladstone09, Heil09}. The low levels of variability seen in current long observations of ULXs with high count rates taken with \xmm~ leaves only one promising source with variability which may be strong enough to detect an rms-flux relation: NGC 5408 X-1. This source has already been shown to display the kinds of timing features often seen in BHBs and AGN including quasi-periodic oscillations and power spectral breaks \citep{Strohmayer07,Strohmayer09}. It has been observed in two separate long \xmm~ ($>$ 100 ks) observations and is relatively bright ($\sim$ 0.8$~ct~s^{-1}$).  In this letter we analyse both observations for evidence of the rms-flux relation. 


\begin{figure*}
\begin{center}$
\begin{array}{ll}
   \includegraphics[width=6.0 cm, angle=90]{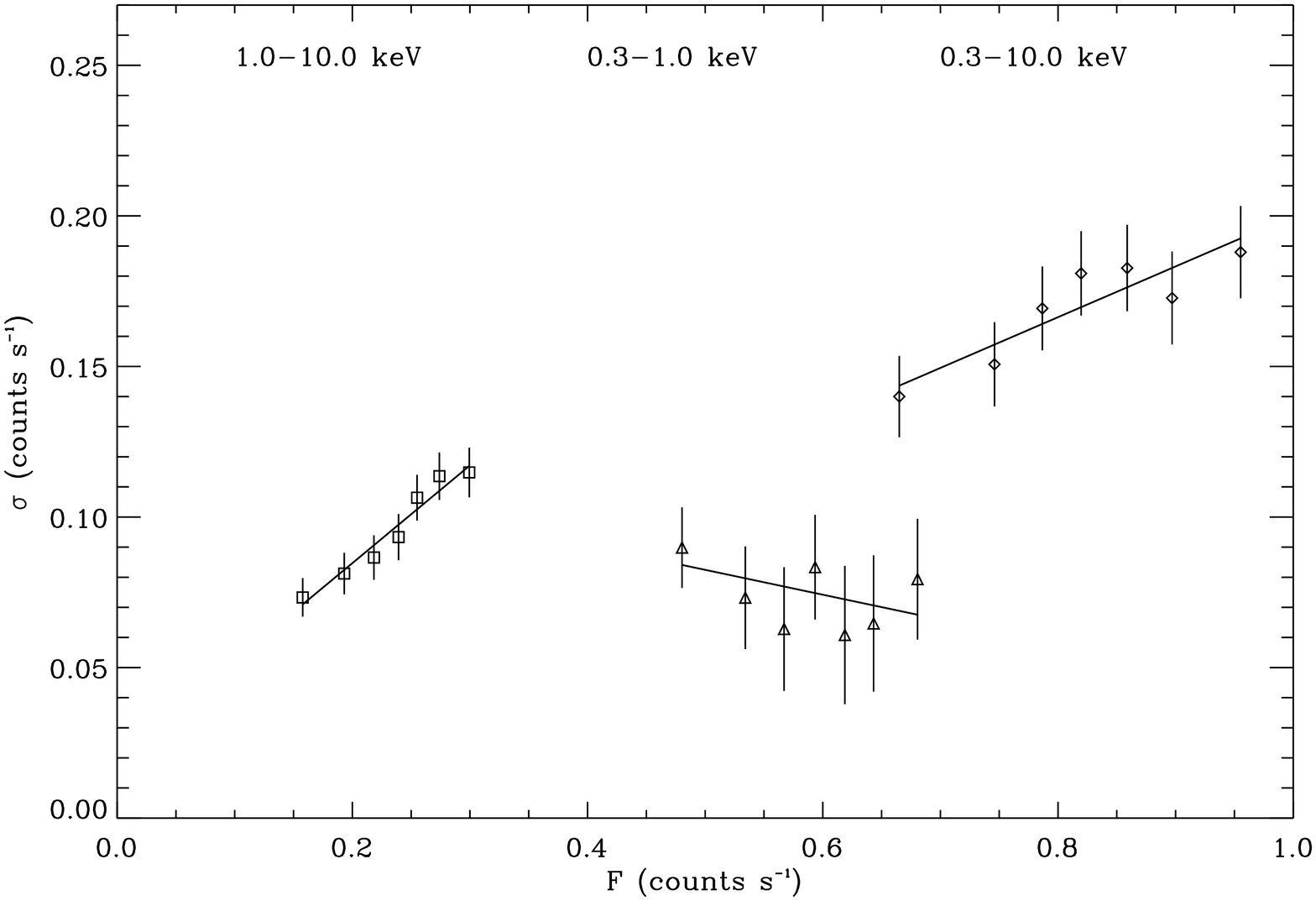} &
	\includegraphics[width=6.0 cm, angle=90]{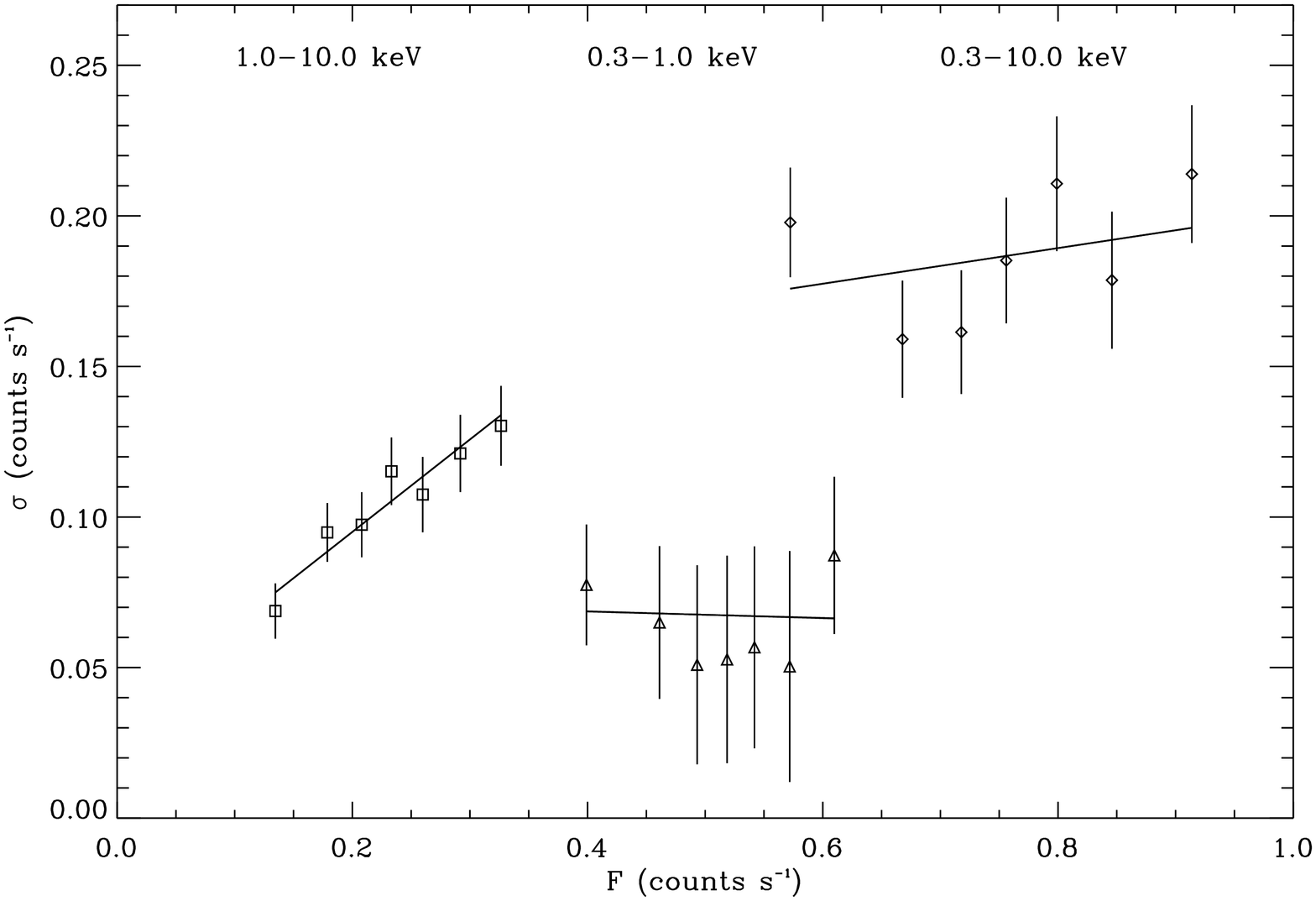}
\end{array}$
\end{center}
\caption{Rms-flux points for each of the three energy bands used in the analysis with 1$\sigma$ errors. \emph{Left}. obs. id. 0302900101; \emph{Right} obs. id. 0500750101.}
\label{fig:energyrms}
\end{figure*}

\section{Data Analysis}
\label{sect:data}

We have used the two longest observations of NGC 5408 X-1 in the \xmm~ archive (Obs. Ids. 0302900101 and 0500750101) taken on the 13th January 2006 and 2008 respectively. The data were extracted from the European Photon Imaging Camera (EPIC) pn camera only, as this provided the highest count rate time series. The pn was operated in full--frame mode for both observations. The data were extracted and reduced following standard procedures using the SAS version 7.1.0. In particular, source events were extracted from a 36 arcsec region centred on the X-ray source and background events from a rectangular region on the same chip as close to the source as allowable. The light curves were initially extracted in the energy band 0.3-10.0 keV and with time resolution of 73.4 ms (the frame time of the camera in this mode).  

	The careful removal of background flares is necessary for good timing analysis. Following the method used in \cite{Heil09} 82.2ks of good time were extracted from the 2006 observation, but only 34.5ks was extracted from the 2008 observation which was particularly affected by background flares. 

Each observation was divided into continuous segments of 150s duration, and for each segment the mean count rate ($\langle F \rangle$ in units of ct s$^{-1}$) and the periodogram was calculated. The periodogram was found using absolute normalisation such that the sum over frequencies multiplied by the frequency resolution ($\Delta\nu$), gives the variance, the square root of which is the rms in absolute units (i.e. also ct s$^{-1}$). The intrinsic randomness of the variability scatters the periodogram points randomly about the underlying power spectrum. In order to suppress these random fluctuations and get a reasonable estimate of the power spectrum as a function of flux the periodograms were averaged in flux bins. The 2006 and 2008 observations were binned such that each flux bin contained 60 and 30 segments, respectively (there are fewer points per bin in the 2008 observation due to the smaller amount of good time). The variance over the $6-50$ mHz range was estimated by summing over the appropriate frequency range in each of the flux averaged periodograms. The Poisson noise level, which is approximately $2\langle F \rangle$ for each periodogram, was subtracted. The square root of the result is an estimate of the intrinsic rms of the source as a function of its flux, $\sigma$. Errors on each rms point were calculated using a more general form of the prescription used by \cite{Gleissner04}, namely propagating the variances of the individual periodogram points as they were averaged into flux bins and summed over the $6-50$ mHz frequency range (full details will be given in Heil et al. 2010 in prep.). This range includes the quasi-periodic oscillations (QPOs) visible in both of these observations -- excluding them would require segments of much longer timescales and hence limit the number of measured rms-flux points. The QPOs are seen at frequencies of 11.4, 19.8 and 27.7 mHz in the 2006 observation \citep{Strohmayer07} and 10.02, 13.5 and 6.0 mHz in 2008 \citep{Strohmayer09}. The analysis was repeated for soft band ($0.3-1$ keV) and hard band ($1-10$ keV) light curves. The resulting rms-flux data are shown in Figure \ref{fig:energyrms}.

\begin{table*}
\centering
\begin{tabular}{lllrrrrrrrr}
\hline\hline

Obs. Id. & Date & Energy &  $\langle F \rangle$ & $N$ & $k$ & $C$ & $\tau$ & $p_{\tau}$ & $\chi^{2}/dof$ & $p_{\chi^2}$ \\
         & & (keV)  & (ct~s$^{-1}$) & & &(ct~s$^{-1}$)&  & & &  \\
(1)      & (2) & (3) & (4) & (5)  & (6)  & (7) & (8) & (9) & (10) & (11) \\

\hline

0302900101 & 2006 Jan 13  & 0.3-10.0 & 0.82 & 7 & 0.17$\pm$0.06 & -0.18$\pm$0.3 & 0.81 & 0.01 & 1.77/5 & 0.87\\
          & & 0.3-1.0  & 0.59 & 7 &-0.08$\pm$0.1 & 1.5$\pm$1.1 & -0.23 & 0.45 & 1.7/5 & 0.88 \\
	  & & 1.0-10.0   & 0.23 & 7 & 0.33$\pm$0.06 & -0.06$\pm$0.05 & 1.0 & 0.001 & 1.46/5 & 0.91 \\
\hline
0500750101 & 2008 Jan 13 & 0.3-10.0 & 0.75 & 7 & 0.06$\pm$0.07 & -2.39$\pm$3.8 & 0.42 & 0.17 & 5.9/5 & 0.31 \\
	   & & 0.3-1.0  & 0.52 & 7 & -0.01$\pm$0.14 & 7.3$\pm$94 & -0.14 & 0.65 & 1.56/5 & 0.90 \\
	   & & 1.0-10.0   & 0.23 & 7 & 0.31$\pm$0.07 & -0.11$\pm$0.07 & 0.91 & 0.004 & 1.99/5 & 0.85 \\
\hline
\multicolumn{7}{c}{{\it }}\\
\end{tabular}
\caption{Results from fitting the rms-flux relation in three separate energy bands errors given are the 1$\sigma$ errors on each value. (1) \xmm~ Observation identifier; (2) Date of observation; (3) Energy band; (4) Mean count rate; (5) Number of points in fitting; (6) Gradient; (7) X-axis intercept; (8) Kendall rank correlation coefficient ($\tau$); (9) $p$-value of $\tau$; (10) $\chi^{2}$ with number of degrees of freedom; (11) $p$-value of $\chi^{2}$ fit.}
\label{table:results}
\end{table*}

\section{Testing for the rms-flux relation}

Following \cite{Gleissner04} we tested the correlation between rms and flux using the Kendall rank correlation coefficient ($\tau$). The results are given in table \ref{table:results} and suggest a correlation is at least present in the 2006 data. Although not highly significant individually the correlation seen in the first and longest observation indicates that an rms-flux correlation could be present. The result for the second observation is less convincing. For comparison with the previous analyses \citep[e.g.][]{Uttley01, Gleissner04} we fitted the resulting points in the full band with a linear function of the form $ \sigma = k(\langle F \rangle - C)$ with $k$ the gradient and $C$ the intercept on the flux axis. The $\chi^{2}$ values from the initial fit were acceptable, with $p$-values of 0.87 and 0.31 for the 2006 and 2008 bands respectively indicating a good agreement with a linear model. 

 \cite{Strohmayer09} noted that below $\sim$1 keV the QPO is no longer visible in the 2008 observation and that the 2006 observation displays a clear difference in the shape of the soft spectrum. For both the 2006 and 2008 observations a lower level of variability was detected in the soft energy band at 12$\%$ and 11$\%$ factional rms respectively, rising to 39$\%$ and 50$\%$ in the hard band. The strong energy dependence within the power spectrum appears to be reflected in the rms-flux relations taken at different energies: we find no strong evidence of a coherent rms-flux relation in the soft energy band,  although a linear relation with an intercept at $C=0.0$ cannot be excluded due to the large uncertainties (caused by the relatively low level of variability). 

\begin{figure}
\begin{center}
   \includegraphics[width=6.0 cm, angle=270]{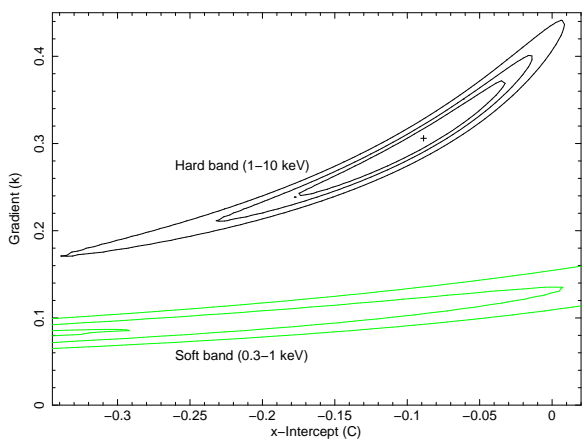} 
\end{center}
\caption{Contour plot of k against C for the hard band ($1-10$ keV, black) and soft band ($0.3-1$ keV, green) rms-flux relations. For each band the 2006 and 2008 observations have been simultaneously fitted. Contours are shown at the $\Delta\chi^{2} = 2.30, 4.61$ and $9.21$ levels which correspond to the 68, 90 and 99 per cent confidence regions for the two parameters.}
\label{fig:cont}
\end{figure}

The parameters of the rms-flux relation in the hard band are broadly consistent between the 2006 and 2008 observations and so we performed a simultaneous fit to the two sets of rms-flux data to obtain tighter constants on the model parameters. The resulting $\chi^{2}$ value was 7.36 with 12 degrees of freedom, indicating the strength of the similarity. Figure \ref{fig:cont} shows the confidence contours for the parameters for this simultaneous fit in each of the two energy bands. The best fitting intercept parameter (i.e the flux at which the rms drops to zero) is indeed negative -- the $99\%$ confidence interval for this one parameter ranges from -0.27 to -0.002 ct s$^{-1}$. As the Figure shows, the parameters for the hard and soft band are significantly different. It must be noted that because the gradient in the soft band is consistent with zero the contours for the soft band extend towards extremely large negative intercepts.

\section{Coherence and time delays}

\begin{figure}
\begin{center}
   \includegraphics[width=7.0 cm, angle=0.0]{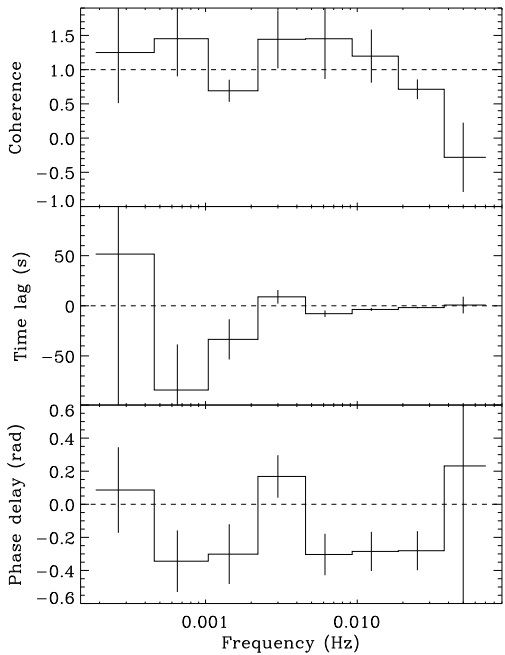} 
\end{center}
\caption{Results of cross-spectral analysis between $0.3-1$ keV (soft band) and
$1-10$ keV (hard band) light curves from the 2006 observation. The top
panels shows the noise-corrected coherence, which is high $(\sim 1$) over
the observable range. The lower two panels show different representations of
the hard-soft delays. The middle panels shows the time delay $\tau(f)$ and
the lowest panel shows the phase delay $\phi(f)$ (which are related by
$\tau(f) = \phi(f)/2 \pi f$). A negative delay means the hard band leads the
soft band. In this case the time delay, or equivalently the phase delay, is
inconsistent with zero delay at all frequencies ($p < 0.002$).}
\label{fig:coh}
\end{figure}

We compared the variability in the two bands ($0.3-1$ keV, and
$1-10$ keV) by computing the coherence and time delay spectra from
the cross-spectrum \citep[see][]{Vaughan97, Nowak99, Vaughan03b}. The coherence was corrected for Poisson noise following the
prescription of \cite{Vaughan97}, and the time delay was computed in
the standard manner \citep[e.g.][]{Nowak99, Vaughan03b}. The
cross-spectral estimates were made using time series binned to $\Delta T =
7.34$ s (i.e. $100$ pn frame times), averaging over segments of length $3.7$
ks ($512$ data points), and over logarithmic frequency bins. The results are shown in Fig.\ref{fig:coh}.

The coherence is consistent with unity, meaning the
variations in one band are well correlated with variations in the other band
(once corrected for the contribution due to Poisson noise in each band). The
delay spectrum is poorly constrained but consistent with a constant phase
delay  ($\phi \sim 0.3$ rad), or a frequency dependent time delay ($\tau
\sim 0.05 f^{-1}$), where the hard band variations precede those in the soft
band. A zero time (or phase) delay model is rejected ($p < 0.002$ in a
chi-square goodness-of-fit test). The shorter 2008 observation was also analysed in a similar manner and is consistent with a phase delay of $\phi \sim 0.2$ rad (a zero phase delay model is rejected with $p = 0.043$). 

\section{Discussion}  

We have shown that the linear rms-flux relation, previously seen in Galactic X-ray binaries and Active Galaxies is also present in the ULX NGC 5408 X-1. This detection is important as it extends the apparent ubiquity of the relation in accreting sources to ULXs. This can be taken as further proof that at least this ULX behaves in a manner similar to other accreting black hole systems. The fact that the rms-flux relation has been observed in the three main types of luminous accreting black hole systems (Galactic Binaries, ULXs and Active Galaxies) suggests that it is either a basic consequence of all luminous accretion onto black holes, or that it only occurs under certain restricted conditions but that these conditions are met for all three types of systems. 

The relation is clearly present above 1 keV, but at softer energies the lower variability amplitude means a linear relation can neither be confirmed nor rejected. However, the coherence of the two light curves indicates a common origin for the variability in the two bands, which would suggest that the soft band light curve may contain a linear rms-flux relation with a very low gradient.

We may attempt to explain the observed rms-flux behaviour, in the most basic manner, in terms of two components behaving in different ways. The first component (A) obeys a linear rms-flux relation with zero intercept and a gradient equal to its fractional rms. The second component (B) is variable but its amplitude is constant with flux. This has the effect of adding constant flux and rms, effectively shifting the origin of the observed relation. The fluxes of components A and B are constrained to be non-negative and the average flux and rms of B can be no higher than the minimum observed flux and rms. The fractional rms of component B can be described by the gradient of the line from the origin to this point. The flux intercept of the linear model ($C$) can be understood in terms of the fractional rms of components A and B; if the fractional rms of B is greater than that of A then the flux intercept will be negative, as observed in the hard-band.

The cross-spectral analysis is also indicative of a time lag which compares favourably to an extrapolation of the lag observed from BHBs to lower frequencies \citep[see e.g.][]{Miyamoto88, Cui97, Nowak99, Pottschmidt00, Miyamoto93}.
Although soft lags have been observed less frequently the majority of observations of BHBs do not reach the low energy range analysed here. \cite{Miyamoto93} have observed soft lags in the very high state of GX 339-4 at low energies (1.2-2.3 keV), which appear similar to those we observe. These
 results, on both the rms-flux relation and time delays, could be confirmed
 with long X-ray observations of other bright, variable ULXs. These further
 observations should help to clarify the explanatory value of our proposed
 variability components (A and B), and how these and the hard-soft X-ray time
 lags relate to the different components thought to explain the energy and
 power spectra of accreting black holes. The present results indicate that
 such intensive studies of ULX variability, although challenging in terms of
 the observational demands, hold great promise for constraining the processes
 driving ULXs behaviour, and the position of ULXs in the scheme of black hole
 accretion from BHBs to AGN.

\section*{acknowledgements}

LMH acknowledges support from an STFC studentship. This paper is based on observations obtained with \xmm, an ESA science mission with instruments and contributions directly funded by ESA Member States and the USA (NASA). We would like to thank the reviewer for their helpful comments and suggestions. 

\bibliographystyle{mn2e}
\bibliography{ULXrms.bib}

\bsp

\label{lastpage}

\end{document}